\documentclass[pre,aps,twocolumn,showpacs]{revtex4-1}
\usepackage{amsmath,amssymb,graphicx}
\usepackage{epsfig}
\usepackage{textcomp}
\usepackage{color}
\usepackage{cases}
\usepackage{epstopdf}
\usepackage{hyperref}
\hypersetup{pdfpagemode=UseNone,colorlinks=true,linktoc=page,pdfborder={0 0 0},linkcolor=blue,citecolor=blue}

\begin{document}
\title{Colonization of a territory by a stochastic population under a strong Allee effect and a low immigration pressure}

\author{Shay Be'er, Michael Assaf, and Baruch Meerson}

\affiliation{Racah Institute of Physics, Hebrew University
of Jerusalem, Jerusalem 91904, Israel}


\begin{abstract}
We study the dynamics of colonization of a territory by a stochastic population at low immigration pressure.
We assume a sufficiently strong Allee effect that introduces, in deterministic theory, a large critical population size for colonization.
At low immigration rates, the average pre-colonization population size is small thus invalidating the WKB approximation to the master equation.
We circumvent this difficulty by deriving an exact zero-flux solution of the master equation and matching it with an approximate non-zero-flux solution of the pertinent Fokker-Planck equation in a small region around the critical population size. This procedure provides an accurate evaluation of the
quasi-stationary probability distribution of population sizes in the pre-colonization state, and of the mean time to colonization, for a wide range of immigration rates. At sufficiently high immigration rates our results agree with WKB results obtained previously. At low immigration rates the results can be very different.

\end{abstract}

\maketitle

\section{INTRODUCTION}
\label{introduction}


Any isolated population, which regulates itself via random births and deaths, is doomed to extinction \cite{Bartlett,Nisbet,OMI}. Large and therefore long-lived stochastic populations ultimately go extinct via a rare sequence of events when random population losses dominate over gains.  This basic extinction scenario, unaccounted for by deterministic theory, is at work in many situations in physics, chemistry, biology and other fields.  One example from epidemiology is extinction of an endemic disease from a population when no new infectives arrive \cite{Bartlett}.  Prior to extinction, a large population resides in a long-lived quasi-stationary state,
with a lifetime (the mean time to extinction) which is exponentially large in the average population size~\cite{OMI}.

It has been long recognized that extinction is prevented by immigration: via either colonization of empty regions, or the ``rescue effect" \cite{Brown,Hanski}. Similarly, arrival of new infected individuals can restart the epidemics in a population which has recovered from an infection. Mathematically, by introducing a constant immigration flux into a stochastic population model, one eliminates the absorbing state at zero population size and therefore prevents extinction.  An important additional effect that many populations exhibit is the Allee effect, by which population biologists mean a group of effects causing a reduction in the per-capita growth rate at small population sizes \cite{Allee}. In the language of deterministic theory, a strong Allee effect introduces a non-zero critical population size for establishment. If there is no immigration, and the initial population size is smaller than the critical size, the population goes extinct quickly. If the initial population size is greater than the critical size, a long-lived state with a large population size appears. We will call this state the colonization state. In the presence of low immigration pressure (by which ecologists mean small immigration rate) the absorbing zero-population state gives way to the pre-colonization state: a state with a small population size. As a result, the population can be either in the pre-colonization state, or the colonization state. The demographic noise (which, for large populations, is weak) causes rare switches between the two states. We will assume that the Allee effect is sufficiently strong so that this critical population size is large, see Sec. \ref{model} for details. Here we evaluate the mean time to colonization (MTC), which we define as the mean switching time between the pre-colonization and colonization states.

Noise-induced switching between long-lived states is a classical paradigm of statistical physics going back to Kramers \cite{Hanggi}.
In the context of population dynamics, describable as a continuous-time Markov process with a discrete space of states,
accurate and useful general expressions for the mean switching time have only become available recently. For single-population systems with single-step processes
(that is, when there are only transitions between a state with $n$ individuals and a state with $n\pm 1$ individuals),
an exact analytical expression for the (properly defined) mean switching time can be obtained ~\cite{Gardiner} by solving a recursive equation for the mean first passage time, see Sec.~\ref{comparison}. This expression, however, is extremely cumbersome and not very informative. Furthermore, for multiple-step processes no exact solutions are available. These difficulties may explain the common practice, especially in the population biology literature \cite{Potapov}, of using the so called ``diffusion approximation". In this approximation the mean switching time is evaluated from a Fokker-Planck equation that is derived from the original master equation via a truncated system-size expansion \cite{vanKampen}.  Unfortunately, the diffusion approximation breaks down in the tails of the quasi-stationary distribution. As shown in many studies \cite{Gaveau,Doering,Assaf2006,KS,spectral,OMI}, this leads to errors in the mean switching time that are exponentially large in the population size, thus invalidating the whole calculation.

A robust and efficient way of evaluating the mean switching time in large populations is provided by a dissipative variant of WKB approximation \cite{Bender}
that employs the average population size as a large parameter and is applied directly
to the original master equation \cite{Kubo}.  In this way Dykman et al \cite{Dykman} calculated the effective entropic barrier that determines the mean switching time up to a pre-exponential factor. More recently, Meerson and Sasorov \cite{explosion} used the WKB formalism to calculate, for a specific model, the mean switching time with account of the pre-exponential factor. This calculation requires going to the sub-leading order of the WKB theory and also dealing with a vicinity of the unstable fixed point where WKB theory breaks down. The approach of Ref. \cite{explosion} was extended to a general set of reactions by Escudero and Kamenev \cite{EK}, see also Ref. \cite{PRE2010}.

The WKB formalism, however, assumes that the average population size in each of the two bistable states is large.
In the colonization problem this assumption breaks down when the immigration pressure is so small that the average
population size in the pre-colonization state is $\mathcal{O}(1)$ or less. Indeed, in some ecological systems the population dynamics represents a series of recurrent extinctions and colonizations (the ``rescue effect") \cite{Brown,Hanski,MacArthur,MacArthurbook}. In this case, and in other, milder, cases \cite{Potapov,low1,low2}, the  immigration rate is very small, and the situation calls for approximations that would respect the non-WKB character of the pre-colonization state. Here we develop an approximate method  that yields the MTC and the long-lived quasi-stationary distribution (QSD) of population sizes in the situation when the pre-colonization state is of a non-WKB nature. The method holds for a broad range of immigration rates. For single-step processes it does not employ the WKB approximation altogether, whereas for multiple-step processes the WKB approximation is only used for sufficiently large $n$ where it is justified.

An important element of our method is the zero-flux solution of the quasi-stationary master equation which can be found by recursion.  For single-step processes, the resulting recursion solution is well known \cite{Gardiner}, and it gives a very good approximation of the quasi-stationary distribution of the pre-colonization state for the population sizes $n$ from $n=0$ to a close vicinity of the Allee threshold (that is, of the unstable fixed point of deterministic theory). For multiple-step processes a recursion solution can often be obtained for sufficiently small $n$~\cite{PRE2010}, and matched with a zero-flux WKB solution that remains valid until close to the Allee threshold.

In the vicinity of the unstable fixed point, and at larger $n$, there is a finite probability flux toward larger $n$ \cite{explosion}. A proper non-zero-flux solution can be found by performing a boundary-layer analysis of the Fokker-Planck equation which can be derived from the original master equation and is valid in the vicinity of the Allee threshold \cite{explosion}. By matching the zero-flux solution with the boundary-layer solution in their joint region of validity, we determine the probability flux and evaluate the MTC.
The resulting MTC exhibits an entropic barrier, so that the MTC is exponentially long.  We show that the WKB approximation remains remarkably accurate at quite low immigration pressures, well beyond conservative estimates. As the immigration pressure decreases, the WKB prediction starts to fail. At not too small immigration rates,  the correct entropic barrier still coincides with that obtained from WKB approximation. There is an important pre-factor, however, that strongly depends on the immigration rate and, for very low immigration pressure, is very different from that predicted by the WKB approximation. At still lower immigration rates even the entropic barrier is different from the WKB prediction. Our result for the  MTC in a broad range of immigration rates is the central result of this work.

For simplicity, we will present our method for a concrete stochastic population model.
Here is a plan of the remainder of the paper. The model is introduced in Sec. \ref{model}. In Sec. \ref{theory} we present a derivation of the QSD of the pre-colonization state and obtain the MTC. In Sec. \ref{comparison} we compare our result for the MTC with the (very cumbersome) exact expression and with a WKB formula. The main results are summarized and discussed in Sec. \ref{discussion_and_summary}, while the Appendix contains a derivation of the Fokker-Planck equation and its approximate solution in the vicinity of the Allee threshold.

\section{MODEL}
\label{model}


We consider a stochastic population describable by a continuous-time and discrete-state Markov process.
When only single-step processes are present, the master equation reads
\begin{equation}
	\label{md-f}
	\frac{dP_n}{dt}=\lambda_{n-1}P_{n-1}-\lambda_nP_n+\mu_{n+1}P_{n+1}-\mu_nP_n,
\end{equation}
where $P_n(t)$ is the probability of observing the population size $n$ ($n=0,1,...$) at time $t$, while $\lambda_n$ and $\mu_n$ are
the effective birth and death rates, respectively.  The deterministic rate equation,
corresponding to the master equation (\ref{md-f}), is
\begin{equation}
	\label{md-b}
	\frac{d\bar{n}}{dt}=\lambda_{\bar{n}}-\mu_{\bar{n}}.
\end{equation}
The specific model \cite{OMII} we will be dealing with incorporates an Allee effect, as modeled by Dennis \cite{Dennis89}, and a steady immigration flux, in a variant of the stochastic Verhulst model \cite{PRE2010,Nasell}. In this model
\begin{equation}
	\label{md-a}
	\lambda_n=r+\frac{Bn^2}{n+N}\quad \mathrm{and} \quad\mu_n=n+\frac{Bn^2}{K},
\end{equation}
where time is rescaled so that the linear term in the death rate is equal to $1$. The effective birth rate $\lambda_n$ accounts for immigration with $n$-independent rate $r$. The $n$-dependent part of $\lambda_n$ is proportional to $n$ at large $n$ and to $n^2$ at small $n$. Together with the linear in $n$ part of the death rate $\mu_n$ this feature accounts for an Allee effect. The coefficient $B\gtrsim 1$ is the reproduction rate. The large parameters $N$ and $K$ control the Allee threshold and the carrying capacity of the colonization state, respectively.

We will work in the parameter regime where Eq.~(\ref{md-b}) with rates~(\ref{md-a}) has three positive fixed points $n_1<n_2<n_3$. The fixed points $n_1$ and $n_3$ are attracting. They correspond, in the deterministic theory, to the pre-colonization and colonization states, respectively. The fixed point $n_2$ is repelling; it determines the Allee threshold. We will assume throughout the paper that the immigration is weak, $r\ll N$. In this regime the fixed point $n_1$  can be obtained by neglecting the nonlinear terms in Eqs.~(\ref{md-a}), while the other two fixed points can be obtained by neglecting the immigration:
\begin{eqnarray}
\label{md-c}
  n_1 &\simeq & r,\nonumber \\
  n_2 &\simeq& \frac{K}{2}\!\left[1-\frac{1}{B}-\frac{N}{K}-\sqrt{\left(1-\frac{1}{B}-\frac{N}{K}\right)^2-\frac{4N}{BK}}\,\right], \nonumber \\
 n_3 &\simeq & \! \frac{K}{2}\!\left[1-\frac{1}{B}-\frac{N}{K}+\sqrt{\left(1-\frac{1}{B}-\frac{N}{K}\right)^2-\frac{4N}{BK}}\,\right]\!.
\end{eqnarray}

\begin{figure}[b]
\includegraphics[width=3.5in,clip=]{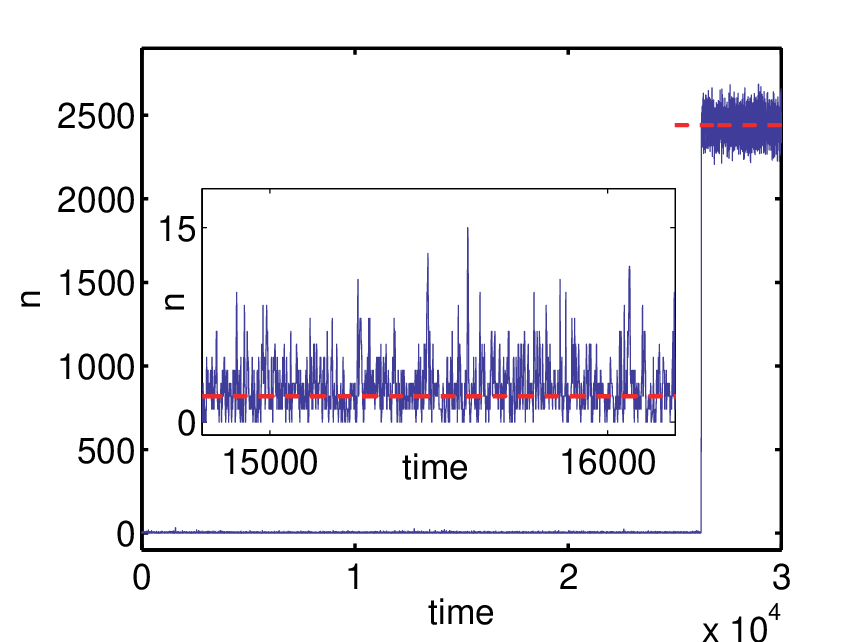}
\caption{(Color online) Solid line: a typical Monte-Carlo realization of the random birth-death process $n(t)$ with the birth and death rates given by  Eq.~(\ref{md-a}). The parameters are $B=2$, $r=2$, $N=30$ and $K=5000$. The initial condition is $n=2$. The dashed line shows the fixed point $n=n_3$ from Eq.~(\ref{md-c}). Inset: a blowup of $n(t)$ in the pre-colonization state. The solid line depicts the simulation, the dashed line shows the fixed point $n=n_1$ from Eq.~(\ref{md-c}).}
\label{fig-a}
\end{figure}

The deterministic equation (\ref{md-b}) ignores demographic noise. The latter causes the population to switch randomly between the pre-colonization and colonization states. Figure~(\ref{fig-a}) shows a typical realization of the stochastic dynamics of the system obtained
in a Monte Carlo simulation employing the Gillespie algorithm \cite{Gillespi} with the rates given in Eq.~(\ref{md-a}).
The initial condition is such that the population finds itself, with probability close to 1, in the pre-colonization state.
One can see that the population dwells over a long time in the pre-colonization state. However, when a rare large fluctuation brings the population over the Allee threshold $n_2$, the population size flows almost deterministically towards the colonization state at $n=n_3$. Our task is to determine the QSD of the pre-colonization state and the MTC.

For simplicity, we will assume that
$K$ is so large that the nonlinear term in the death rate $\mu_n$ is negligible \cite{largeK}.  With the new rates,
\begin{equation}
	\label{md-d}
	\lambda_n=r+\frac{Bn^2}{n+N}\quad \mathrm{and} \quad\mu_n=n,
\end{equation}
the fixed point $n=n_3$ moves to infinity, and the switching problem is replaced by an effective problem of noise-driven population explosion.
In this problem the population size, once it overcomes the Allee threshold, blows up in a finite time. This time scale is of deterministic nature and therefore relatively short \cite{explosion}. The fixed points $n_1$ and $n_2$ become
\begin{eqnarray}
	\label{md-e}
  \frac{n_1}{N} &=& \frac{r}{N}\left[1+\mathcal{O}\left(\frac{r}{N}\right)\right],\quad \mathrm{and} \nonumber \\
  \frac{n_2}{N}&=& \frac{1}{B-1}\left[1-B\frac{r}{N}+\mathcal{O}\left(\frac{r^2}{N^2}\right)\right],
\end{eqnarray}
respectively.

\section{SOLUTION}
\label{theory}


\subsection{Recursive solution}
\label{RS}


When starting from a sub-threshold initial condition, the stochastic population relaxes, with a high probability, to the pre-colonization state around the stable fixed point $n_1$ of the deterministic theory. Although long-lived, this state is metastable,
as there is a nonzero probability flux through the unstable fixed point $n_2$ towards large $n$.  At times much longer than the deterministic relaxation time (let us call it $t_r$),  the pre-colonization probability distribution is described by the eigenvector of the master equation with the smallest positive eigenvalue $1/\tau$, where $\tau$ is the MTC \cite{KS,explosion,EK,PRE2010}:
\begin{equation}\label{QSDansatz}
P_n(t)\simeq \pi_n e^{-t/\tau}.
\end{equation}
Here $\pi_n$ is the QSD of the pre-colonization state. Plugging Eq.~(\ref{QSDansatz}) into Eq.~(\ref{md-f}) and neglecting the exponentially small term $-\pi_n/\tau$ on the left hand side, one arrives at a stationary difference equation for the QSD \cite{KS,explosion,EK,PRE2010}:
\begin{equation}\label{difference}
\lambda_{n-1}\pi_{n-1}-\lambda_n \pi_n+\mu_{n+1}\pi_{n+1}-\mu_n \pi_n =0.
\end{equation}
The disregard of the term $-\pi_n/\tau$ can only be justified if the immigration rate $r$ is much larger than
$1/\tau$: a criterion that can be checked a posteriori.

Equation~(\ref{difference}) is exactly soluble via recursion \cite{Gardiner}, and the zero-flux solution has the form
\begin{equation}\label{recgen}
\pi_n = \pi_0\prod^{n-1}_{k=0} \frac{\lambda_k}{\mu_{k+1}},
\end{equation}
where $\pi_0$ is determined from normalizing the total probability to one \cite{normalization}.
For  $\lambda_k$ and $\mu_k$  from Eq.~(\ref{md-d}) we obtain,
with a help of ``Mathematica",
\begin{equation}\label{RS-a}
\pi_n  = \frac{\pi_0 B^{n-1} rN!}{n!(n+N-1)!} \frac{\Biggl|\Gamma\left[n+\frac{r}{2B}+i\frac{\sqrt{r(4BN-r)}}{2B}\right]\Biggr|^2}
{\Biggl|\Gamma \left[1+\frac{r}{2B}+i\frac{\sqrt{r(4BN-r)}}{2B}\right]\Biggl|^2},
\end{equation}
where $\Gamma(\dots)$ is the gamma function. This solution is valid for all $n$ from zero to a close vicinity
of the unstable fixed point $n_2$, see Refs.~\cite{explosion,EK,PRE2010} and the next section. Employing the weak immigration assumption (that we have assumed in Sec. \ref{model}), that is, $r\ll N$, the pre-colonization QSD is sharply peaked around $n_1\simeq r$. As a result, we can neglect, for the purpose of normalization, the second term, proportional to $B$, in the birth rate~(\ref{md-d}). The remaining simple immigration-death process is described by the Poisson distribution with mean $r$, and we find the normalization constant $\pi_0\simeq e^{-r}$.

The distribution tail is described by the full expression~(\ref{RS-a}) that we will now simplify using the strong inequalities $n\sim N\gg 1$. Let us introduce the rescaled population size $q=n/N$ that can be treated as a continuous variable. We use the Stirling formula $k! \simeq \sqrt{2\pi} \,k^{k+1/2}e^{-k}$ for the factorials of Eq.~(\ref{RS-a}). For  the squared absolute value of the gamma function in the numerator of Eq.~(\ref{RS-a}) we can write $|\Gamma(a+ib)|^2=\Gamma(a+ib) \Gamma(a-ib)$. Using the Stirling formula in each of the multipliers, we obtain
$$
|\Gamma(a+ib)|^2 \simeq 2\pi e^{-2a}(a^2+b^2)^{a-1/2}e^{-2b \arctan (b/a)}.
$$
After some algebra, all this yields
\begin{eqnarray}\label{RS-c}
\pi(q)&\simeq&\frac{rA\sqrt{2\pi(q+1)}}{B\sqrt{Nq^3}}\nonumber\\
&\times& e^{N\left[q\ln\left(\frac{Bq}{q+1}\right)-\ln(q+1)\right]-\frac{r}{B}\left[B+\frac{1}{q}-\ln(N q)\right]},
\end{eqnarray}
where
\begin{equation}
	\label{RS-b}
A=\Biggl|\Gamma\left(1+\frac{r}{2B}+i \,\sqrt{\frac{N r}{B}-\frac{r^2}{4B^2}}\right)\Biggr|^{-2}.
\end{equation}
We will also need a more specialized asymptotic of Eq.~(\ref{RS-c}) in a close vicinity of the unstable fixed point: $N^{-1/2}\ll q_2-q\ll 1$, where $q_2=n_2/N$. Here it suffices to expand the logarithm of $\pi(q)$ in powers of $q-q_2$ up to second order:
$\ln \left [ \pi(q) \right ] = a_0+a_1(q_2-q)+a_2(q_2-q)^2$. As the coefficients $a_1$ and $a_2$ are multiplied by a factor proportional to $q-q_2\ll 1$, it suffices to calculate them only in the leading order in $N\gg 1$. The coefficient $a_0$ demands a higher accuracy, and contributions of the order of ${\cal O}(N)$, ${\cal O}(N^{1/2})$ and ${\cal O}(1)$ need to be kept. By doing so, and keeping terms up to  ${\cal O}(1)$ in the exponent, we can approximate Eq.~(\ref{RS-c}) as
\begin{eqnarray}
	\label{RS-d}
	&&\hspace{-3mm} \pi(q)\simeq \frac{\sqrt{2\pi}rA(B-1)}{\sqrt{BN}}\nonumber \\
&&\hspace{-3mm} \times e^{-N\left\{\ln\left(\frac{B}{B-1}\right)-\frac{r}{BN}\left[2B-1+\ln\left(\frac{B-1}
{N}\right)\right]+\frac{(B-1)^2}{2B}(q_2-q)^2\right\}}.
\end{eqnarray}

\subsection{Boundary-layer solution}
\label{BL}


The zero-flux approximation for the QSD, found in the previous section, is invalid close to the unstable fixed point $n_2$ and at larger $n$, where the proper solution has a non-zero flux \cite{explosion}. A major simplifying factor here is the validity of the  Fokker-Planck approximation in a narrow boundary layer  around $n_2$:  $|n-n_2|\ll n_2$ or $|q-q_2|\ll 1$ (to remind the reader, we assume $B\gtrsim 1$). The Fokker-Planck approximation is valid here because  the QSD varies sufficiently slowly with $n$: $|(\pi_{n+1}-\pi_n)/\pi_n|\ll 1$~\cite{explosion,EK,PRE2010}. The derivation of the Fokker-Planck equation and its solution in the boundary layer was presented elsewhere~\cite{explosion,EK,PRE2010}. For the reader's convenience, we briefly reproduce these calculations in the Appendix. The boundary layer solution reads
\begin{equation}
	\label{BL-a}
	\pi^{(BL)}(q)=\frac{\sqrt{\pi}\tilde{J}_c}{\ell} \,e^{\frac{(q_2-q)^2}{\ell^2}}\mathrm{erfc}\left(\frac{q-q_2}{\ell}\right),
\end{equation}
where
\begin{equation}\label{ell}
\ell^2 = \frac{1}{N} \frac{\lambda(q_2)+\mu(q_2)}{\lambda^{\prime}(q_2)-\mu^{\prime}(q_2)},\quad \tilde{J}_c = \frac{J_c}{\lambda^{\prime}(q_2)-\mu^{\prime}(q_2)}.
\end{equation}
Here primes denote the derivatives with respect to the argument, $\lambda(q)=\lambda_n/N$ and $\mu(q)=\mu_n/N$. $J_c$ is the a priori unknown constant probability flux through the unstable fixed point $n=n_2$. We will now determine it by matching the boundary layer solution~(\ref{BL-a}) with the ``bulk solution", that is the zero-flux recursive solution (\ref{RS-a}), in their joint region of validity $N^{-1/2}\ll q_2-q\ll 1$. The bulk solution is described in this region by the asymptotic (\ref{RS-d}). Now we approximate the boundary layer solution~(\ref{BL-a}) in this region. As $\ell\sim N^{-1/2}$, we can use the asymptotic $\text{erfc}\,(-z) \simeq 2$ at $z\gg 1$. Then, using Eqs. (\ref{md-d}) and (\ref{md-e}), we obtain after some algebra
\begin{equation}
	\label{BL-b}
	\pi^{(BL)}(q)\simeq \sqrt{2\pi BN}J_c \, e^{\frac{(B-1)^2N}{2B}(q_2-q)^2}.
\end{equation}

\subsection{Quasistationary distribution and mean time to colonization}
\label{MTC}

\begin{figure}[t]
\includegraphics[width=3.0in,clip=]{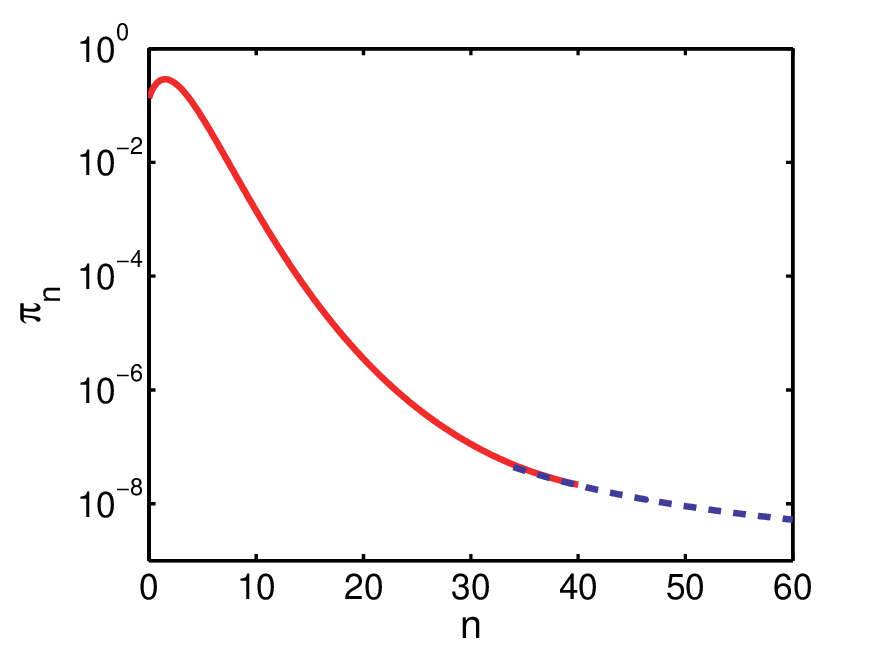}
\caption{(Color online) The QSD $\pi_n$ as a function of $n$ for $B=2$, $r=2$ and $N=50$ in a log scale. The QSD includes two overlapping  asymptotics: the bulk asymptotic, given by Eq.~(\ref{RS-a}) with $\pi_0=e^{-r}$ (solid line), and the boundary layer asymptotic, given by Eq.~(\ref{BL-a}) with the flux $J_c$ from Eq.~(\ref{MTC-a}) (dashed line).}
\label{fig-b}
\end{figure}

Demanding that the expressions~(\ref{RS-d}) and~(\ref{BL-b}) coincide, we determine
the probability flux $J_c$:
\begin{equation}
	\label{MTC-a}
	J_c \simeq A\frac{(B-1)r}{BN}e^{-N\left\{\ln\left(\frac{B}{B-1}\right)-\frac{r}{BN}\left[2B-1+\ln\left(\frac{B-1}{N}\right)\right]\right\}}.
\end{equation}
The QSD is now fully determined. The bulk of the QSD, for $0\leq n\lesssim n_2$, is given by the zero-flux recursive solution Eq.~(\ref{RS-a}) with $\pi_0=e^{-r}$.  In the boundary layer $|n-n_2|\ll n_2$ the QSD is given by Eq.~(\ref{BL-a}) with $J_c$ from Eq.~(\ref{MTC-a}) and $A$ from Eq.~(\ref{RS-b}). Figure~(\ref{fig-b}) shows a plot of the QSD for a specific choice of parameters.

We are now in a position to determine the MTC. Let us return to Eq.~(\ref{md-f}) for the time-dependent probability $P_n(t)$
and sum it over $n$ from $n=0$ to $n=n_2$. As $P_n(t)$, at times $t\gg t_r$, is described by Eq.~(\ref{QSDansatz}),
the left hand side becomes
\begin{equation}\label{lhs}
\sum_{n=0}^{n_2} \frac{dP_n}{dt} \simeq - \frac{1}{\tau} \,e^{-t/\tau} \sum_{n=0}^{n_2} \pi_n \simeq - \frac{1}{\tau}\,e^{-t/\tau},
\end{equation}
where we have used the fact that $\pi_n$ is normalized to $1$, and the normalization is mostly contributed to by relatively small $n$'s. The summation over the right hand side of Eq.~(\ref{md-f}) can be split into two parts:
\begin{equation}\label{rhs1}
 \sum_{n=0}^{n_2} (\dots) = \sum_{n=0}^{n_-}(\dots) +\sum_{n=n_-}^{n_2} (\dots),
\end{equation}
where $n_-<n_2$. Let us choose $n_-$ so that it satisfies the double strong inequality $N^{1/2}\ll n_2-n_-\ll N$, or $N^{-1/2}\ll q_2-q_-\ll 1$, where $q_-=n_-/N$. Because of the inequality $N^{1/2}\ll n_2-n_-$ the probability flux at any $n<n_-$ is approximately zero, see the previous subsection. As a result, the first sum on the right hand side of Eq.~(\ref{rhs1}) is zero. In its turn, the inequality $n_2-n_-\ll N$ guarantees the applicability of the the Fokker-Planck approximation on the interval $n_-<n<n_2$.  Therefore, the second term on the right in Eq.~(\ref{rhs1}) can be approximated as
\begin{equation}\label{rhs2}
    - N \,e^{-t/\tau} \int_{q_-}^{q_2} \frac{dj(q)}{dq}\,dq,
\end{equation}
where $j(q)$ is the probability flux in the boundary layer. Taking the integral in Eq.~(\ref{rhs2}), we obtain
\begin{equation}\label{rhs3}
   - N \left[j(q_2)-j(q_-)\right]\,e^{-t/\tau}.
\end{equation}
We choose $q_-$ to satisfy $q_2-q_-\gg N^{-1/2}$, so that $j(q_-)\simeq 0$. In its turn, $j(q_2)\simeq J_c$. Putting
it all together, we obtain the MTC:
\begin{equation}
	\label{MTC-b}	 \tau \simeq \frac{1}{NJ_c}=\frac{B}{A(B-1)r} e^{N\left\{\ln\left(\frac{B}{B-1}\right)+
\frac{r}{BN}\left[2B-1+\ln\left(\frac{B-1}{N}\right)\right]\right\}}.
\end{equation}
This expression, valid in a broad range of immigration rates $r$, is the central result of this paper. It is instructive  to consider different limits when this expression can be simplified. They are determined by the parameter $\gamma=\sqrt{Nr/B}$ that enters
Eq.~(\ref{RS-b}) for $A$. We obtain
\begin{subnumcases}{\label{MTC-c} A^{-1}\simeq}
1,&  $\gamma\ll 1$, \label{MTC-c-c}\\
|\Gamma(1+i\gamma)|^2, & $\gamma=\mathcal{O}(1)$, \label{MTC-c-b}\\
2\pi\gamma \,e^{\frac{r}{B}\ln \gamma -\pi\gamma}, & $\gamma\gg 1$. \quad\quad\quad\quad\label{MTC-c-a}
\end{subnumcases}
\noindent
The corresponding asymptotics of the MTC are
\begin{subnumcases}{\label{MTC-d} \tau\simeq}
\frac{B}{r(B-1)} \,e^{N\ln\left(\frac{B}{B-1}\right)}, & $r\ll 1/N$, \label{MTC-d-a}
\\ \frac{|\Gamma(1+i\gamma)|^2B}{r(B-1)}e^{N\ln\left(\frac{B}{B-1}\right)}, & $r=\mathcal{O}(1/N)$, \label{MTC-d-b}
\\ \frac{2\pi\gamma B}{r(B-1)}e^{N\ln \left ( \frac{B}{B-1} \right )-\pi\gamma} & \nonumber
\\ \times e^{\frac{r}{B}\left\{2B-1+\ln\left[\frac{\gamma(B-1)}{N}\right]\right\}}, & $r\gg 1/N$. \quad\quad\quad\quad\label{MTC-d-c}
\end{subnumcases}

\section{Comparison with the exact and WKB results}
\label{comparison}
\subsection{Exact solution}
\label{exact}


\begin{figure}[b]
\includegraphics[width=3.0in,clip=]{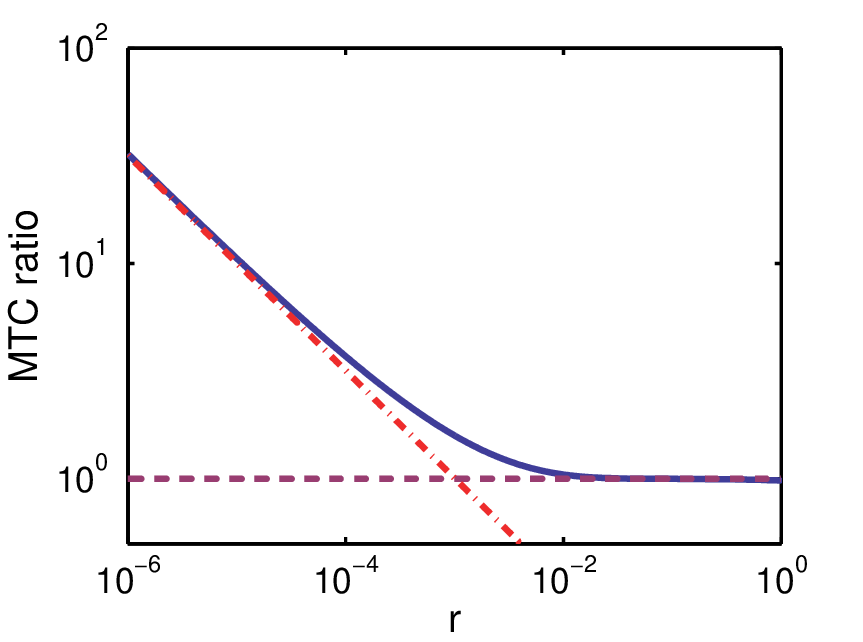}
\caption{(Color online) The ratio of the MTCs as a function of $r$ for $B=2$ and $N=50$ in a log-log scale. Dashed line: $\tau/\tau^{exact}$, where $\tau$ is given by Eq.~(\ref{MTC-b}) and $\tau^{exact}$ is given by Eq.~(\ref{com-a}). Solid line: $\tau/\tau^{WKB}$, where $\tau^{WKB}$ is given by the asymptotic limit $r\ll N$  of Eq.~(23) of Ref.~\cite{EK}. Dashed-dotted line: our theoretical prediction of $\tau/\tau^{WKB}=1/(2\pi\gamma)$ in the limit of $\gamma\ll 1$, where $\tau|_{\gamma\ll 1}$ is given by Eq.~(\ref{MTC-d-a}) and $\tau^{WKB}|_{\gamma\ll 1}$ is the $\gamma\ll 1$ asymptotic of Eq.~(23) of Ref.~\cite{EK}.}
\label{fig-c}
\end{figure}

As we already mentioned, for single step processes the (properly defined) MTC can be found exactly from the backward master equation \cite{Gardiner}. Although quite cumbersome, the exact solution is useful for our purposes, as it enables us to test the accuracy of our approximate result, Eq.~(\ref{MTC-b}).

\begin{figure}[t]
\includegraphics[width=3.8 in,clip=]{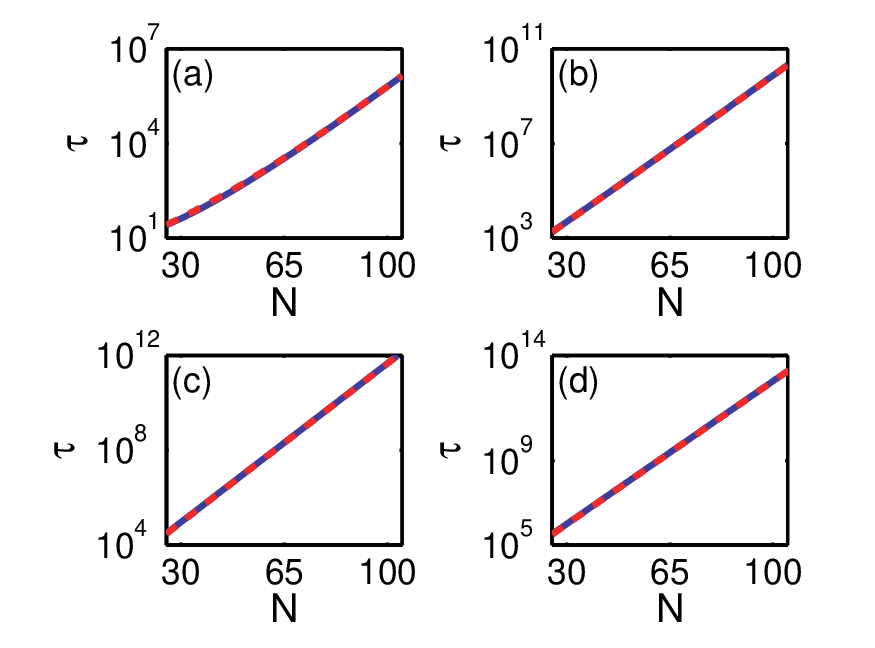}
\caption{(Color online) The MTC as a function of $N$ in a log scale for $B=5$ and $r=1$, $0.1$, $0.01$, and $0.001$ in (a), (b), (c), and (d), respectively. Solid line: $\tau$ from Eq.~(\ref{MTC-b}). Dashed line: $\tau^{exact}$ from Eq.~(\ref{com-a}).}
\label{fig-d}
\end{figure}

The exact derivation supposes that a single-step stochastic process with the birth rate $\lambda_n$ and death rate $\mu_n$ is confined to the interval $a\leq n\leq b$, where $a$ and $b$ are the reflecting and absorbing boundaries, respectively. The exact solution depends on the initial value $n$ of the stochastic process. The mean time  $\tau(n)$ for the process to be absorbed at $n=b$ obeys the exact equation \cite{Gardiner}
\begin{equation}
	\label{com-aa}
	\lambda_n\left[\tau(n+1)-\tau(n)\right]+\mu_n\left[\tau(n-1)-\tau(n)\right]=-1,
\end{equation}
which should be solved with the boundary conditions
$$
\tau(a-1)=\tau(a)\quad \mathrm{and}\quad \tau(b)=0.
$$
In our case, $\lambda_n$ and $\mu_n$ are taken from Eq.~(\ref{md-a}), while $a=0$. A reasonable choice of $b$ is the closest integer to the colonization fixed point $n_3=K(1-1/B)$, whereas the initial number of individuals $i$ is set to be the closest integer to $r$.
The solution is ~\cite{Gardiner}
\begin{equation}
	\label{com-a}
	\tau^{exact}(i)=\sum_{j=i}^{n_3}\phi(j)\sum_{k=0}^j1/\left[\lambda_k\phi(k)\right],
\end{equation}
where
$$
\phi(k)=\prod^k_{l=1}\frac{\mu_l}{\lambda_l}.
$$
Although the products and sums in Eq.~(\ref{com-a}) can be brought to hypergeometric functions, it is more practical to evaluate Eq.~(\ref{com-a}) numerically. Figures~\ref{fig-c} and \ref{fig-d} show comparisons of the exact result with
that predicted by Eq.~(\ref{MTC-b}) at different $r$ and $N$. As one can see, excellent agreement is observed for all relevant values of parameters. We also checked that the agreement is insensitive to the choice of the initial value $n$ in the exact solution, as long as $n<n_2$ and sufficiently far from $n_2$.

We also compared Eq.~(\ref{MTC-b}) with results of extensive Monte Carlo simulations (not shown) and found excellent agreement.

\subsection{WKB approximation}
\label{WKB}

It is assumed in the existing formulations of the WKB theory that all relevant fixed points scale with the population size~\cite{Kubo,Dykman,explosion,EK,PRE2010}. In practice,  the WKB approximation is expected to hold in our colonization problem as long as the pre-colonization fixed point corresponds to a sufficiently large population, even if $r\ll N$. It is interesting to find out how large the pre-colonization population should be for the WKB theory to be accurate. We achieved this goal by comparing
our approximate result for the MTC with that obtained via WKB approximation.

To calculate the MTC in the WKB approximation, $\tau^{WKB}$, in the limit of $r\ll N$, we used Eq.~(23) of Ref. \cite{EK} with the rates given by Eq.~(\ref{md-d}). We took the rates in the leading WKB order.  As $r$ dominates the birth rate in the vicinity of the pre-colonization fixed point, it has to be included in the leading WKB order. We simplified the result by employing the smallness of the parameter $r/N$.   These calculations show that $\tau^{WKB}$ coincides with Eq.~(\ref{MTC-d-c}) in the limit of $r\gg 1/N$. This inequality is much weaker than the na\"{\i}vely expected condition $r\gg1$ that would guarantee that the pre-colonization fixed point $n_1\simeq r$ is describable by a deterministic theory. It is surprising that the WKB theory remains accurate at much lower immigration pressures than one could have expected~\cite{WKB}.

For $r\lesssim 1/N$, the WKB approximation breaks down. This is clearly seen in Fig.~\ref{fig-c}, where we compare our result~(\ref{MTC-b}) for the MTC with the exact result and with the WKB result. In the limit of very low immigration pressure, $r\ll 1/N$, there is a large factor missed by the WKB theory. This factor can be written as
\begin{equation}
	\label{com-b}
	\frac{\tau}{\tau^{WKB}}=\frac{1}{2\pi\gamma}=\frac{1}{2\pi}\sqrt{\frac{B}{Nr}},\;\;\;\;r\ll 1/N.
\end{equation}
For $r$ that is \emph{exponentially} small in the parameter $N$, the factor (\ref{com-b}) describes an effective increase of the entropic barrier to colonization, thus invalidating the WKB approximation in its entirety \cite{nottoosmall}.

\section{SUMMARY AND DISCUSSION}
\label{discussion_and_summary}


We have investigated the dynamics of colonization of a territory by a stochastic population at low immigration pressure. Against all odds, and regardless of how small the immigration rate is, demographic noise eventually drives the population to the colonization state via a rare fluctuation that allows the population to overcome the Allee threshold.

The specific model \cite{OMII} we have dealt with incorporates an Allee effect, as modeled by Dennis \cite{Dennis89}, and a steady immigration flux, in a variant of the stochastic Verhulst model \cite{PRE2010,Nasell}. We have determined the quasi-stationary distribution (QSD) of the population sizes and the mean time to colonization (MTC) in a broad range of immigration pressures. In all parameter regions, our result for the MTC is in excellent agreement with the exact result and with Monte Carlo simulations. At moderate and high immigration rates our results agree with the previously found WKB results. At low immigration rates we obtain a large preexponential correction to the WKB result, due to the breakdown of the latter at low immigration rates. The correction factor becomes huge, and invalidates the WKB result completely, for very low immigration rates.

The calculation method that we have presented here is free of uncontrolled assumptions and can be used for a broad class of stochastic population models that exhibit an Allee effect and low colonization pressure. It can also be extended to multi-step processes that, in general, do not admit exact solutions. In those cases the master equation can be linearized at small population sizes in order to determine the recursive solution there. The recursive solution can be matched, at $n\gg 1$, with a bulk solution obtained by using the (leading and subleading order of the) WKB approximation \cite{explosion,EK,PRE2010}. The WKB approximation breaks down in the vicinity of the Allee threshold. There one has to match the WKB solution with the universal boundary-layer solution, obtained by solving the pertinent Fokker-Planck equation in a close vicinity of the Allee threshold. This double matching procedure is applicable for a broad class of multi-step processes involving arbitrary immigration rates, and it yields the QSD in the entire region of interest, and the MTC.

In this work we have focused on colonization due to demographic stochasticity. It would be interesting to investigate the
interplay of demographic and environmental stochasticity \cite{KamMS,LM,Aetal,AMR} in the colonization under an Allee effect.
It would be also interesting to study colonization under an Allee effect when new immigrants arrive in groups rather than separately \cite{OMII,inprep}.

\section*{Acknowledgments}

This work was supported by Grant No. 300/14 of the Israel Science Foundation and by Grant No. 2012145 from the U.S.-Israel Binational Science Foundation (BSF).


\section*{Appendix}
\label{Appendix}
\renewcommand{\theequation}{A\arabic{equation}}
\setcounter{equation}{0}
We briefly present here a derivation and solution of the quasi-stationary Fokker-Planck equation in the vicinity of the deterministic Allee threshold $n=n_2$, see
Sec.~\ref{model}. Let us define $\lambda(q)=\lambda_n/N$ and $\mu(q)=\mu_n/N$, where $\lambda_n$ and $\mu_n$ are the effective birth and death rates, respectively, given by Eq.~(\ref{md-d}). In terms of the rescaled variable $q=n/N$ the quasi-stationary master equation (\ref{difference}) becomes
\begin{align}
	\label{Ap-a}
	& \lambda(q-1/N)\pi(q-1/N)-\lambda(q)\pi(q) \nonumber \\ &\quad +\mu(q+1/N)\pi(q+1/N)-\mu(q)\pi(q)=0.
\end{align}
Let us denote $f_{+}(q)=\lambda(q)\pi(q)$ and $f_{-}(q)=\mu(q)\pi(q)$. Taylor-expanding $f_{\pm}(q \mp 1/N)$ around $q$ we find
\begin{equation}
	\label{Ap-a2}
	f_{\pm}(q \mp 1/N)\simeq f_{\pm}(q) \mp f_{\pm}^{\prime}(q)/N + f_{\pm}^{\prime\prime}(q)/(2N^2).
\end{equation}
Plugging Eq.~(\ref{Ap-a2}) into (\ref{Ap-a}) we arrive at the quasistationary Fokker-Planck equation
\begin{equation}
	\label{Ap-b}
	[f_{+}^{\prime}(q)-f_{-}^{\prime}(q)]-\frac{1}{2N}[f_{+}^{\prime\prime}(q)+f_{-}^{\prime\prime}(q)]=0,
\end{equation}
that can be written as $\partial_q J(q)=0$, where $J$ is the probability flux. Thus, the solution to this differential equation is a constant-flux solution. In order to proceed we notice that $\pi^{'}(q) \sim N\pi(q)$. Therefore, $f_{+}^{'}(q)$ is governed in the leading order by $\lambda(q)\pi^{'}(q)$, while $f_{-}^{'}(q)$ is governed by $\mu(q)\pi^{'}(q)$.
Integration over $q$  yields, in the leading order of $1/N\ll 1$,
\begin{equation}
	\label{Ap-c}
	[\lambda(q)-\mu(q)]\pi(q)-\frac{1}{2N}[\lambda(q)+\mu(q)]\pi^{\prime}(q)=J_c,
\end{equation}
where $J_c$ is the constant probability flux through the unstable fixed point. We can simplify Eq.~(\ref{Ap-c}) in the boundary layer $|q_2-q|\ll 1$ (see Sec.~\ref{BL} for the definition of the boundary layer) by expanding the drift term up to the first order in $q-q_2$ and putting $q=q_2$ in the diffusion term. The resulting equation accepts the universal form~\cite{explosion,EK,PRE2010}
\begin{equation}
	\label{Ap-d}
	(q-q_2)\pi(q)-\frac{\ell^2}{2}\pi^{\prime}(q)=\tilde{J}_c,
\end{equation}
where $\ell$ and $\tilde{J}_c$ are defined in Eq.~(\ref{ell}) of the main text. Solving the linear first-order Eq.~(\ref{Ap-d}), we obtain
\begin{equation}
	\label{Ap-d2} \pi^{(BL)}(q)=C e^{\frac{(q_2-q)^2}{\ell^2}}
+\frac{\sqrt{\pi}\tilde{J}_c}{\ell}e^{\frac{(q_2-q)^2}{\ell^2}}\mathrm{erf}\left(\frac{q_2-q}{\ell}\right),
\end{equation}
where $C$ and $\tilde{J}_c$ are constants yet to be determined. $C$ can be found by demanding that the boundary layer solution behaves properly at $q>q_2$ \cite{explosion}. Indeed, by considering the asymptotic of the solution at $q-q_2\gg \ell$ we find that $C=\sqrt{\pi}\tilde{J}_c/\ell$ in order to eliminate a rapid exponential growth. As a result, the boundary layer solution takes the form of Eq.~(\ref{BL-a}) of the main text.



\begin{thebibliography}{99}
\bibitem{largeK} As we checked, the criterion for neglecting the nonlinear term in the death rate $\mu_n$ is $\sqrt{K}\gg N$.
\bibitem{WKB} The technical reason for the good accuracy of the WKB asymptotic at $r\gg N^{-1}$ is the rapid convergence of the factor $A$ from Eq.~(\ref{RS-b}) to its leading-order Stirling approximation at $r\gg N^{-1}$.
\bibitem{nottoosmall} To remind the reader, our formalism, based on the calculation of the QSD, demands that $r$ be still much larger than the $1/\tau$.
\end{thebibliography}
\end{document}